\documentclass[fleqn,10pt]{wlscirep}
\usepackage[utf8]{inputenc}
\usepackage[T1]{fontenc}
\usepackage{amsfonts, amsmath, amssymb}
\usepackage{amsthm}
\usepackage{graphicx}
\usepackage{physics}
\usepackage[normalem]{ulem}
\usepackage{xcolor}
\usepackage{caption}
\usepackage{tikz}
\usetikzlibrary{quantikz2}
\usepackage{float}
\usepackage{verbatim}
\usepackage{hyperref}
\usepackage[capitalise, nameinlink]{cleveref}
\usepackage{booktabs}
\usepackage{multirow}
\usepackage{nicematrix}
\usepackage{orcidlink}
\usepackage{braket}
\usepackage{tabularx}
\usepackage{doi}

\usepackage{cancel}

\theoremstyle{plain}

\title{Experimental implementation of a discrete-time quantum walk on biological networks}

\author[1,*]{Viacheslav Dubovitskii}
\author[2]{Filippo Utro}
\author[2]{Aritra Bose}
\author[2]{Laxmi Parida}
\author[1]{Sabrina Maniscalco}
\author[1]{Sergey N. Filippov}

\affil[1]{Algorithmiq, Via della Chiusa 15, 20123 Milan, Italy}
\affil[2]{IBM Quantum, Yorktown Heights, New York, NY, USA}

\affil[*]{Corresponding author: viacheslav.dubovitskii@algorithmiq.fi}

\begin{abstract}
Quantum walks provide a versatile framework for probing the structural and dynamical properties of complex systems ranging from biological networks to synthetic materials. However, their realization on current noisy pre-fault-tolerant quantum computers is fundamentally limited by decoherence. Conventional dense encodings of graph structures require prohibitively deep circuits, making them incompatible with existing hardware. Here we introduce an algorithm that leverages symmetry-sector encoding and trades circuit depth for the number of qubits, while integrating symmetry-respecting postselection as an effective noise-mitigation strategy. This combination enables us to execute practical quantum-walk circuits for biological network motifs on actual quantum hardware. We benchmark the proposed methodology against known state-of-the-art circuit architectures, highlighting significant reduction of circuit depth in our approach at the cost of moderate qubit overhead. Utilizing 40 qubits, we implement quantum walks on complex graphs containing up to 17 nodes and 20 edges---the largest experiment on superconducting hardware to date, with the Hellinger fidelity exceeding 87\% throughout 7 steps. We present a proof‑of‑principle case study that illustrates how experimentally obtained quantum-walk dynamics on a protein-protein-interaction network can be applied to prioritizing disease-associated genes. We discuss the framework scalability in the pre-fault-tolerant era and its potential for studying larger biological networks.

\end{abstract}

\begin{document}

\flushbottom
\maketitle
\thispagestyle{empty}

\section*{Introduction}

Network medicine offers a systematic approach to understanding complex disease mechanisms by examining interactions within biological networks, a cornerstone of precision medicine. In this context, quantum walks (QWs), which leverage quantum principles such as superposition and interference, have emerged as a promising method for modeling information spreading on graphs. Their appeal lies in the ability to capture richer propagation dynamics and explore network topology more deeply than classical approaches~\cite{Bose2026, Kempe_2003, reitzner2011quantum, venegas2012quantum, biamonte2019complex, chen2025hybrid}. Recent studies have demonstrated the potential of QWs in addressing key challenges in network medicine, including link prediction, disease target identification, network propagation, and node classification~\cite{maniscalco2022quantumnetworkmedicinerethinking, e25050730, Saarinen2024, dubovitskii2025quantum, park2025advancing}.

Among the different QW formulations~\cite{reitzner2011quantum,venegas2012quantum,biamonte2019complex,chen2025hybrid}, discrete‑time quantum walks (DTQWs) have seen significant theoretical progress toward implementation on present-day digital quantum processors. However, despite advances in circuit design, their practical implementation on current quantum computers has been constrained by hardware limitations such as gate and measurement errors, leakage noise, and restricted qubit connectivity~\cite{corcoles-2020, fauseweh_quantum_2024, Cai2023-cy, filippov2024scalabilityquantumerrormitigation, Eisert2025-qi, Zimboras2025-me}.

Earlier work on DTQW implementation, conducted during periods of qubit scarcity, prioritized minimizing the required number of physical qubits~\cite{Acasiete2020-tm, Koch2020-ld, Shakeel2020-hq, Georgopoulos2021-uu}. Most circuit-based DTQW implementations used $\lceil\log_2 N\rceil$ qubits (dense encoding for $N$ walker positions) plus a coin qubit, realizing increment-decrement update rules~\cite{Shakeel2020-hq, razzoli2024efficient, Wing-Bocanegra2023, Georgopoulos2021-uu}. Ancilla‑free and coinless constructions pushed this reduction even further. While these approaches enabled early experimental demonstrations, they often did so at the cost of substantially increased multi‑controlled gate complexity leading to deeper circuits that typically degrade overall fidelity~\cite{Wing-Bocanegra2023}. Consequently, few implementations targeting multiple walk steps consistently reported improved performance when multi-controlled gates were reduced or avoided altogether~\cite{Wing-Bocanegra2023, Wadhia2024-nb}. To mitigate the associated resource overhead, several gate-level optimizations have been proposed~\cite{Shakeel2020-hq, razzoli2024efficient, Singh2021-gf, nandi2025robust, Saha2022-ya}; however, despite these efforts, the resulting circuits remain practically viable only for small graph instances.

As qubit quality improved, larger DTQW experiments became possible~\cite{razzoli2024efficient,Zhou2020-ee, Wadhia2024-nb}. However, hardware-informed classical simulations indicated that executing larger DTQWs (e.g., on a 16-node graph over 16 steps) still required significantly lower noise levels to maintain acceptable fidelity~\cite{Zhou2020-ee, Wadhia2024-nb}. For instance, an experiment with 4-qubit circuit (8-vertex cycle graph) achieved fidelity $0.8$ for up to 13 steps, whereas an experiment with 3-qubit circuit (4-vertex cycle graph) reached the fidelity nearly $0.9$ for up to $19$ steps. The latter experiment benefits from the specific structure of the 4-vertex cycle graph that allowed the circuit depth to be almost independent of the number of time steps~\cite{razzoli2024efficient}, a property that does not generalize to other graphs. From a resource perspective, dense encoding remains impractical as decoherence degrades superposition and entanglement in deep circuits. More advanced encoding schemes address this by applying position-dependent coin operators in parallel by introducing ancillary position and coin wires, trading circuit depth for ancilla count~\cite{Nzongani2024-iz} and illustrating a deliberate transition toward using more qubits to enable deeper parallelism and control at the expense of ancilla overhead.

The previously discussed DTQW implementations employ approaches designed for regular graphs (e.g., lines, cycles, lattices) and were validated primarily on these topologies, which limits generalization to irregular or heterogeneous networks such as those arising in biological settings~\cite{Wing-Bocanegra2023, razzoli2024efficient, Wadhia2024-nb}. In response, in order to extend DTQW circuits to irregular and complex networks Refs.~\cite{Mukai2020-fj, Singh_2021, sato2024, sato2025coinedquantumwalkscomplex} deal with an encoding in which position and coin registers explicitly represent node ($i$) and outgoing‑edge labels ($i\rightarrow j$). While structurally transparent, the design of Ref.~\cite{Singh_2021} enables a simple SWAP-based shift; however, it requires coin operations conditioned on node-specific bit patterns, typically through ancilla-assisted checks. As graph structure becomes more heterogeneous, this results in rapidly increasing circuit depth and makes the approach difficult to scale beyond highly regular topologies. Ref.~\cite{sato2024} allocates $\lceil\log_2 N\rceil$ qubits to node positions and $\lceil\log_2 |E|\rceil$ qubits to $|E|$ edge labels, whereas recent Ref.~\cite{sato2025coinedquantumwalkscomplex} exploits $2 \lceil\log_2 N\rceil$ qubits in a dual-register encoding. Functionality of these approaches is validated on 4- and 8-node complex networks via classical simulation~\cite{sato2024} and experiment~\cite{sato2025coinedquantumwalkscomplex}, though their dependence on deep multi-controlled gate structures limits its robustness under realistic noise models. Across these developments, advanced error-suppression strategies are rarely employed, leaving fidelity to depend on circuit shallowness rather than algorithmic techniques. 

Several works exploit encodings within symmetry sectors of the Hilbert space such as single-photon time-bin states that are robust against noise~\cite{Chang2024-ln, Chang2025-xf}. In the context of quantum computing, Cai’s review highlights a simple yet effective mitigation approach: identify violations of the symmetries of the ideal quantum state and remove them via postselection~\cite{Cai2023-cy}. However, making symmetry checks feasible on hardware requires an encoding that exposes measurable symmetries and avoids excessive multi-controlled gates that amplify noise.

Motivated by this need, we introduce an alternative encoding of DTQW states that trades circuit depth for qubit count (twice the number of edges, $2|E|$), a feasible approach on currently available large-scale quantum computers. By encoding information in a specific symmetry sector of the Hilbert space, we significantly reduce circuit depth, while symmetry-respecting postselection enables efficient noise mitigation. It is the combination of reduced circuit depth and the postselection in the proposed algorithm that makes the DTQW less susceptible to noise, leading to the largest DTQW implementation on complex graphs to date ($40$ qubits and a depth exceeding $200$ entangling layers), in contrast to the regular or symmetric structures explored previously~\cite{razzoli2024efficient, wing2025circuit, nandi2025robust}. Our design handles graphs with the nonuniform degree distributions typical of biological networks. We implement the DTQW experimentally on IBM’s quantum hardware with heavy-hex connectivity and show how the resulting propagation dynamics enables the disease-target identification. While universal scalability remains a long-term objective, our design introduces a distinct paradigm by prioritizing hardware-efficient mapping, providing a practical path toward maximizing the utility of current quantum computers.

The paper is organized as follows. The Results section presents our framework for implementing DTQWs on arbitrary complex networks using current quantum computers, including the proposed encoding and postselection strategies, and demonstrates its feasibility through practical hardware experiments on three biologically relevant graphs. We then compare our encoding approach with other baselines in terms of the circuit size, depth, and accuracy, yielding insights into circuit scalability and optimization. A case study illustrates how experimentally obtained DTQW dynamics can be applied to prioritizing disease genes. The Discussion section contextualizes these findings and outlines future directions, while Methods detail the DTQW formulation and the procedures used for performance evaluation and quantum-walk-based node prioritization.

\section*{Results}
\label{sec:results}
\subsection*{Implementation framework}

We present a novel framework for implementing DTQW on arbitrary complex networks. The framework is designed to be compatible with digital gate-based quantum processing units (QPUs). Consider a network with $N$ nodes $V=\{1,2,\dots,N\}$ and undirected edges $E=\{\{i,j\}\mid i \text{~and~} j \text{~are~adjacent}\}$. The DTQW Hilbert space is spanned by the directed-edge states $\ket{i\rightarrow j}$ that represent a walker at node $i$ which is about to hop to adjacent node $j$~\cite{Mukai2020-fj}. The Hilbert space dimension is $2|E|$ and coincides with the number of directed edges $(i,j) \in E_{\rm dir}$. The quantum state $\ket{\psi(t)}$ of the DTQW at step $t$ is defined in the abstract $2|E|$-dimensional Hilbert space as
\begin{equation}
    \label{eq:state}
    \ket{\psi(t)} = \sum_{(i,j)\in E_{\rm dir}}\psi_{ij}(t)\ket{i\rightarrow j}.
\end{equation}
The probability of finding the walker at node $i$ is
\begin{equation}
    \label{eq:probability}
    P_i(t) = \sum_{j:\, (i,j)\in E_{\rm dir}} \big|\braket{i\rightarrow j| \psi(t)}\big|^2 = \sum_{j:\, (i,j)\in E_{\rm dir}} |\psi_{ij}(t)|^2.
\end{equation}

\begin{figure}[H]
\centering
\includegraphics[width=0.85\textwidth]{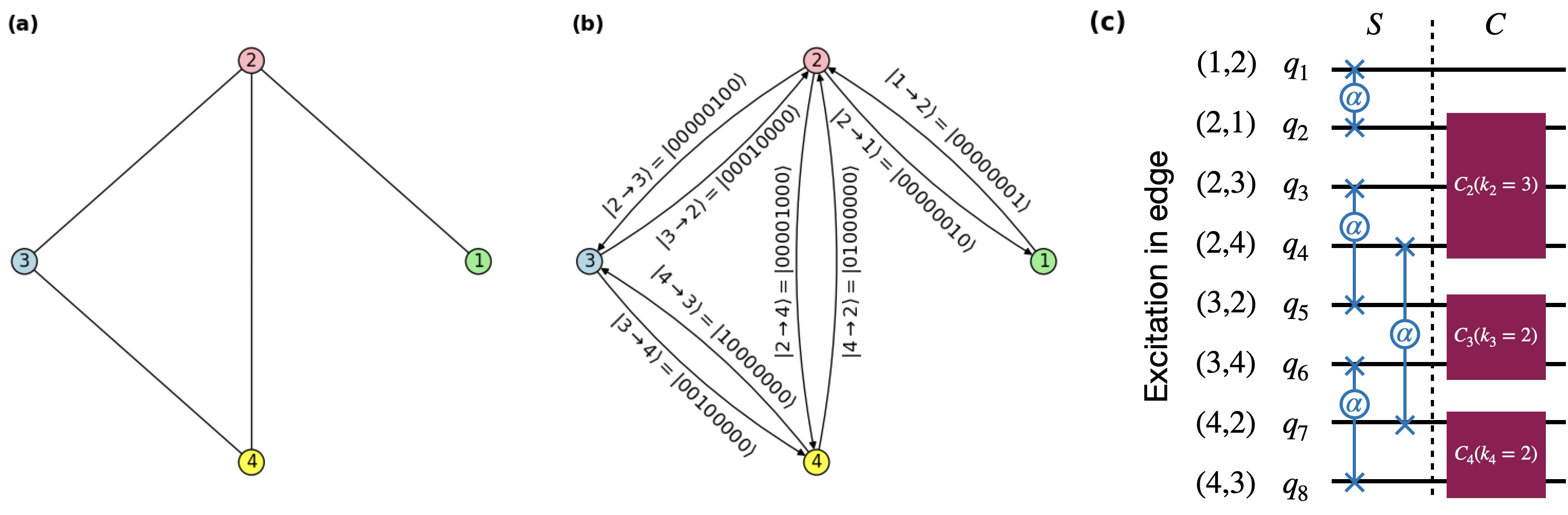}
\caption{\textbf{Implementation framework.}
\small (a) Example 4-node graph. Nodes 1, 2, 3, and 4 have degree 1, 3, 2, and 2, respectively. (b) Encoding DTQW states $\ket{i \rightarrow j}$ into qubit states. The length of qubit register is twice the number of graph edges. (c) Single-step in DTQW circuit consisting of the shift and coin operators. The shift operator $S$ represents a collection of partial-swap gates associated with opposite directions within the same edge, whereas the coin operator $C$ is composed of the smaller Grover coins $C_i$ associated with the graph nodes $i$ of degree $k_i$. Qubits carry information about excitations (walker) in the corresponding graph edges.} \label{fig:implementation} 
\end{figure}

The structure and complexity of our implementation are determined by the encoding of states $\ket{i \rightarrow j}$ into the qubit space. Specifically, we represent the DTQW state $\ket{\psi(t)}$ through computational basis states restricted to the single-excitation subspace~\cite{geller_universal_2015}, $\ket{0\dots 001}, \ket{0\dots 010},\dots \ket{1\dots 000}$, i.e., the states of Hamming weight 1 also known as the bracelet states. Hence, in the target qubit space, the walker's position and direction is represented by basis states where only one qubit is in the excited state $\ket{1}$. We motivate our choice by the presence of energy loss that makes basis states with higher Hamming weight less stable over time as qubits typically tend to decay from the excited state $\ket 1$ to the ground state $\ket 0$. This approach requires as many qubits as double-counting of graph edges, $2|E|$, and exploits the single-excitation symmetry for robustness of noise-reduction algorithms. We illustrate the encoding approach on a small graph consisting of four edges shown in Fig.~\ref{fig:implementation}.

We prepare the initial state $\ket{\psi(0)}$ as a uniform superposition of the states residing on a specific node $i_0$ of interest (seed in a biological network),
\begin{equation}
    \label{eq:state0}
    \ket{\psi(0)} = \frac{1}{\sqrt{k_{i_0}}}\sum_{j:\, (i_0,j)\in E_{\rm dir}}\ket{i_0 \rightarrow j},
\end{equation}
where $k_{i_0}$ is the degree of node $i_0$. In our encoding, the initial state is entangled and represents a $W$-state~\cite{Cabello2002-mo} whose preparation on a quantum computer is both straightforward and efficient.

We consider the DTQW with the evolution operator $U = (CS)^t$ for step $t$, where $C$ ($S$) is a coin (shift) operator. The coin operator is of a Grover type, because its phases are invariant under the permutation of neighboring nodes $\lbrace j_1, j_2, \dots, j_{k_i} \rbrace$~\cite{Prabhu_Influence} (see details in Methods). 
Importantly, we introduce two innovations. First, in order to achieve an efficient decomposition of the quantum circuit into native gates on a hardware with limited connectivity (e.g., heavy-hex), we label the states $\ket{i\rightarrow j}$ resting on the same node $i$ with the neighboring bracelet states to keep the minimal Hamming distance in the qubit gate operations within the coin operator. Second, in the considered biological networks, where nodes have degrees $1$, $2$, and $3$, the depth of the coin-operator circuit is additionally reduced by means of the appropriate basis rotation (see Methods).

In contrast to the shift operator realized via the swap gates~\cite{Mukai2020-fj} that effectively flips the state $\ket{i\rightarrow j}$ to $\ket{j\rightarrow i}$, our formulation generalizes this structure by introducing a tunable parameter $\alpha$, which  
models quantum interference between forward and backward transitions:
\begin{equation}
    \label{eq:shift}
    S\ket{i\rightarrow j} = \sqrt{\alpha}\ket{j \rightarrow i} + \text{i} \sqrt{1-\alpha} \ket{i \rightarrow j}.
\end{equation}
The case $\alpha = 0$ corresponds to no transitions, whereas $\alpha = 1$ realizes a swap operation that does not increase the number of interfering terms (the operation is non-entangling in the qubit space) and introduces an imbalance between odd and even evolution steps. We therefore choose $\alpha = 1/2$ corresponding to an $\sqrt{\text{iSWAP}}$ operation that creates maximal coherence in the subspace spanned by $\ket{i\rightarrow j}$ and $\ket{j\rightarrow i}$ and is maximally entangling in the qubit space. 

To find a probability distribution \eqref{eq:probability} of the walker at step $t$, we compose the circuit as follows: an initial-state preparation \eqref{eq:state0}, followed by alternating blocks of the shift and coin operators (a total of $2t$ blocks), and finally measurement in the computational basis. The circuit is executed multiple times to ensure statistically meaningful outcomes. 

Due to the presence of noise in both quantum gates and measurements, the experimental design is intentionally built to make allowance for that: only measurement outcomes with Hamming weight equal to 1 are retained for computing the probability \eqref{eq:probability}. It is the postselection of measurement bitstrings from the single-excitation subspace that serves as an effective noise filter and substantially improves fidelity. After postselection, the retained measurement outcomes are aggregated according to graph connectivity to yield the node-level probabilities \eqref{eq:probability}. The resulting probabilities for progressively increased time steps $t$ constitute the basis for downstream biological data analysis.

\subsection*{Hardware experiment on biological data}

To evaluate the practical applicability of the DTQW circuit design described above, we applied it to three graphs derived from real-world biological data. The construction of these graphs, along with the procedures for qubit mapping, circuit optimization, and QPU data processing, is described in Methods. Following the proposed framework, we built quantum circuits corresponding to 7 DTQW steps for each graph. The characteristics of the graphs and the corresponding circuits are summarized in Table~\ref{Tab:graphs}.

\begin{table}[ht]
\centering
\caption{Structural properties of biological networks, corresponding quantum circuits, and performance indicators.} \label{Tab:graphs}
\renewcommand{\arraystretch}{0.95}
\setlength{\tabcolsep}{5pt}
\small
\begin{tabular}{c c c c c | c c c c c c c c c | c c}
\cline{1-16}
\multicolumn{5}{c|}{\textbf{Biological graph}} & \multicolumn{9}{c|}{\textbf{Transpiled quantum circuit}} & \multicolumn{2}{c}{\textbf{Performance}} \\
\cline{1-16}
 & \textbf{Nodes} & \textbf{Edges} & & & & \multicolumn{7}{c}{\textbf{Entangling layers across time steps}} & \textbf{CZ gates} & \multicolumn{2}{c}{\textbf{Fidelity}} \\
\cline{7-13}
\textbf{\#} & $N$ & $|E|$ & \textbf{Diameter} & \textbf{Density} & \textbf{Qubits} 
 & \textbf{1} & \textbf{2} & \textbf{3} & \textbf{4} & \textbf{5} & \textbf{6} & \textbf{7} & \textbf{per step} & \textbf{$F_{\rm H}$} & \textbf{$F_{\rm HBC}$} \\
\cline{1-16}
1 & 11 & 12 & 6 & 0.22 & 24 & 36 & 66 & 96 & 126 & 157 & 191 & 223 & 165 & $\geq 0.95$ & $\geq 0.71$ \\
2 & 15 & 18 & 8 & 0.17 & 36 & 52 & 86 & 120 & 154 & 188 & 231 & 264 & 286 & $\geq 0.9$ & $\geq 0.46$ \\
3 & 17 & 20 & 8 & 0.15 & 40 & 45 & 85 & 125 & 165 & 205 & 245 & 287 & 322 & $\geq 0.87$ & $\geq 0.54$ \\
\cline{1-16}
\end{tabular}
\end{table}

\begin{figure}[ht]
\centering
\includegraphics[width=0.87\textwidth]{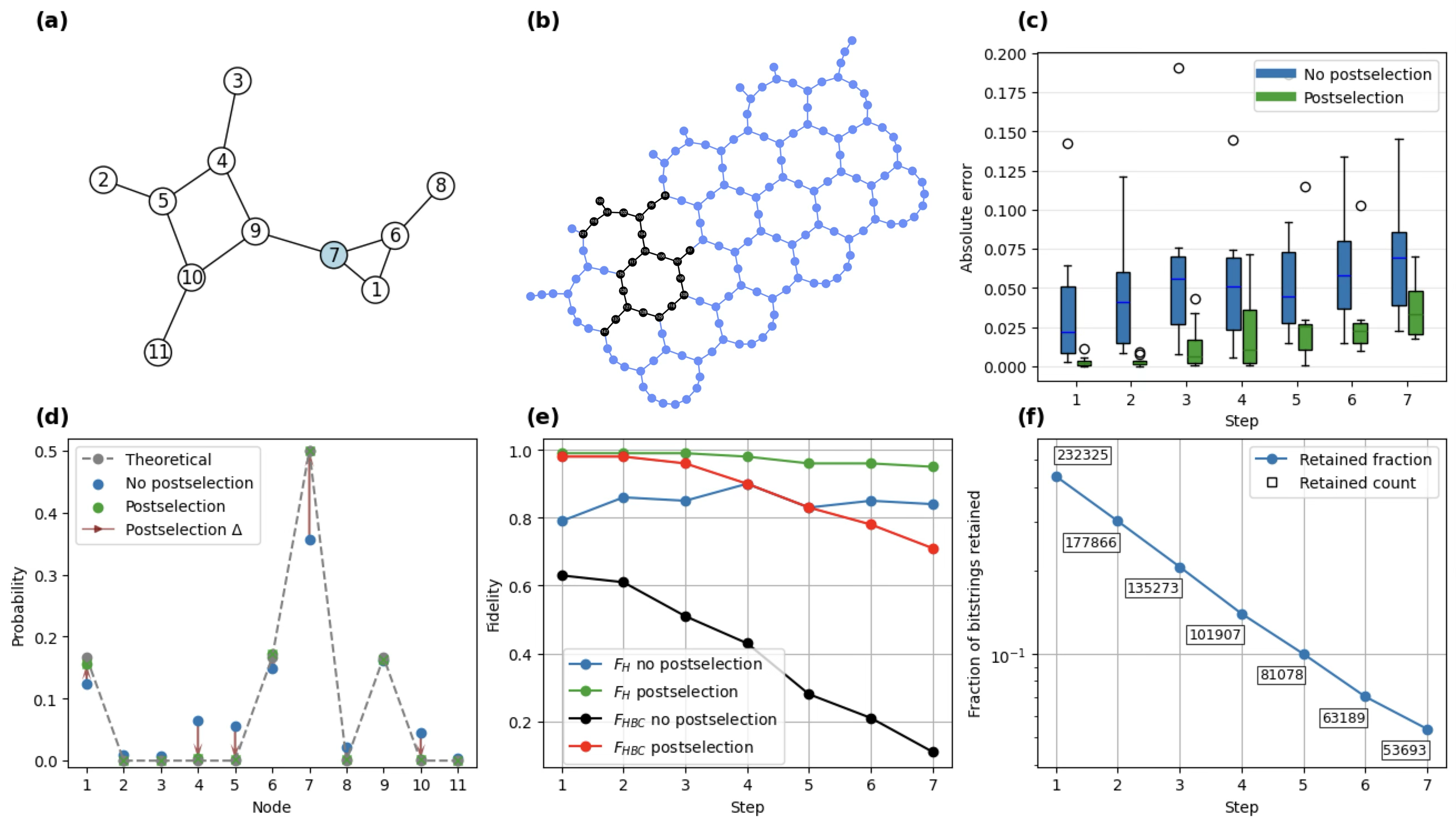}
\caption{\textbf{Quantum walk on an 11-node, 12-edge graph (24-qubit encoding).} 
\small (a) Subgraph extracted from the BioPlex 3.0 PPI network. Nodes 1--11 correspond to genes PON2, VAMP5, MTOR, LYPD3, FLVCR1, HLA-C, HLA-DQA1, HLA-G, TMEM214, FAM234B, and ADPGK, respectively. Quantum walker is initialized at the highlighted node. (b) 24-qubit circuit layout for the experiment performed on \texttt{ibm\_kingston} with heavy-hex connectivity. (c) Boxplot of absolute errors between experimental and simulated probabilities across nodes per step. (d) Probability distribution at Step~1 . Brown vertical arrows show the change between paired probabilities introduced by postselection for each node. (e) Fidelity with and without postselection. (f) Postselection ratio and shot count across steps.}
\label{fig:probs_4}
\end{figure}

\begin{figure}[ht]
\centering
\includegraphics[width=0.87\textwidth]{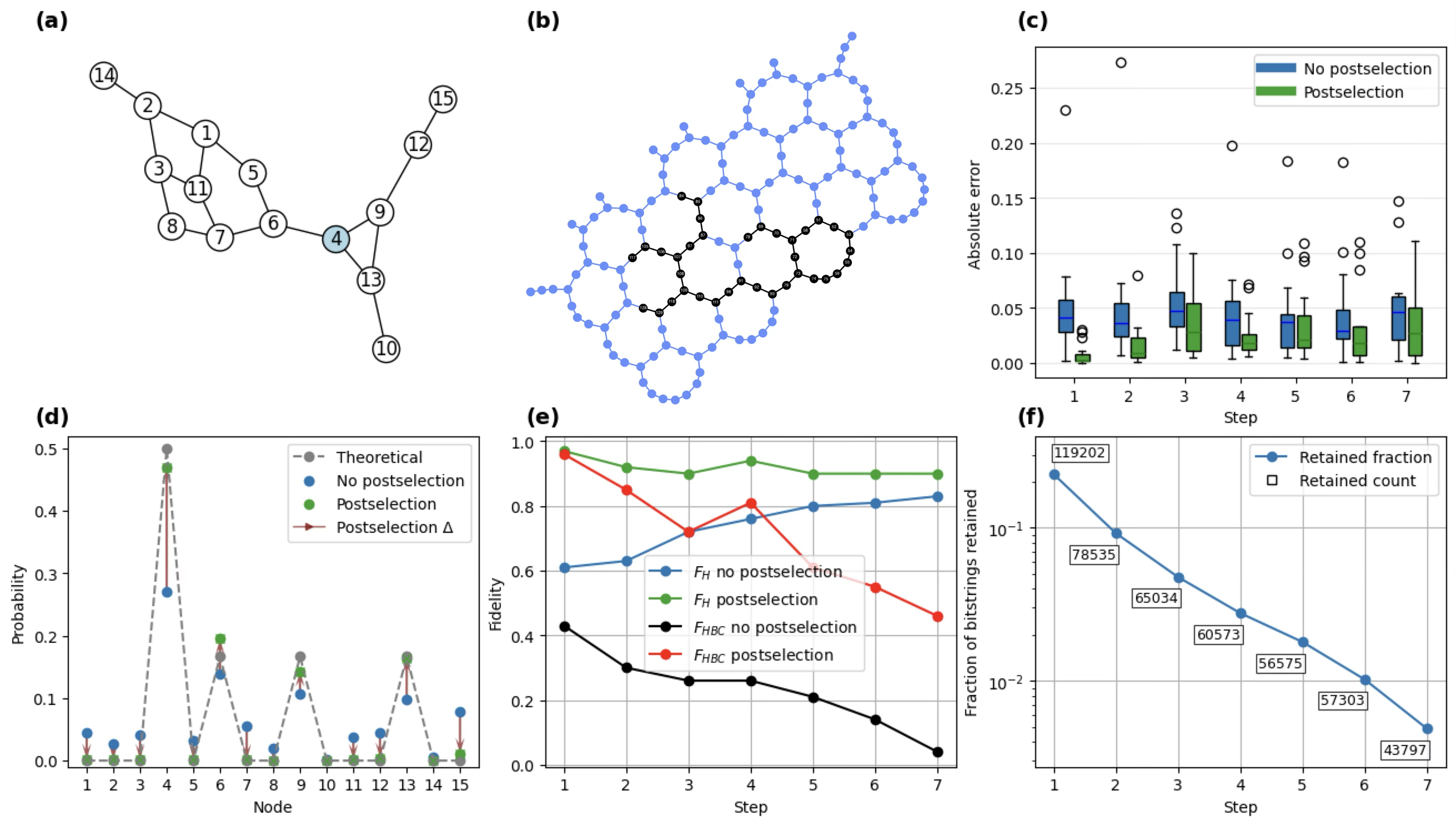}
\caption{\textbf{Quantum walk on a 15-node, 18-edge graph (36-qubit encoding).} 
\small (a) Subgraph extracted from the BioPlex 3.0 protein–protein interaction network. Nodes 1--15 correspond to genes IGF2BP3, APOBEC3D, APOBEC3F, SF3B1, TRUB2, ABT1, ZFR, RBMS2, SNRPA1, SNRPB2, LIN28A, COIL, PHF5A, YTHDC1, SART3, respectively. Quantum walker is initialized at the highlighted node. (b) 36-qubit circuit layout for the experiment performed on \texttt{ibm\_pittsburgh} with heavy-hex connectivity. (c) Boxplot of absolute errors between experimental and simulated probabilities across nodes per step. (d) Probability distribution at Step~1 . Brown vertical arrows show the change between paired probabilities introduced by postselection for each node. (e) Fidelity with and without postselection. (f) Postselection ratio and shot count across steps.}
\label{fig:probs_10}
\end{figure}

\begin{figure}[ht]
\centering
\includegraphics[width=0.87\textwidth]{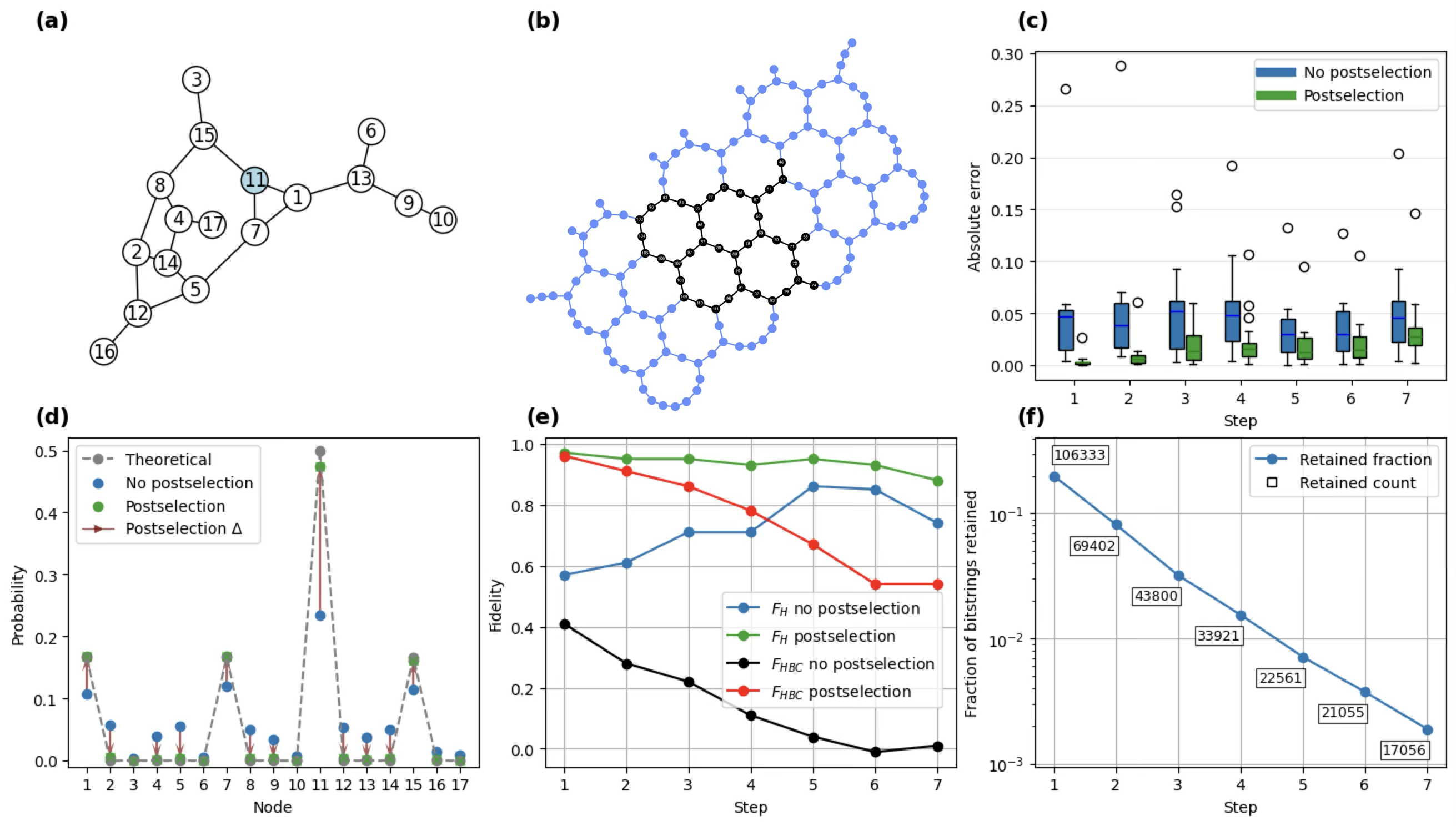}
\caption{\textbf{Quantum walk on a 17-node, 20-edge graph (40-qubit encoding).} 
\small (a) Subgraph extracted from the BioPlex 3.0 PPI network. Nodes 1--17 correspond to genes AHCYL1, ATF7, AGAP3, ATF2, EEF1D, GPC4, AHCYL2, FOSL2, GPC1, HADHA, NFKBIL1, BATF3, CAMKV, FOSL1, KBTBD7, RTL6, CREB5, respectively. Quantum walker is initialized at the highlighted node. (b) 40-qubit circuit layout for the experiment performed on \texttt{ibm\_kingston} with heavy-hex connectivity. (c) Boxplot of absolute errors between experimental and simulated probabilities across nodes per step. (d) Probability distribution at Step~1 . Brown vertical arrows show the change between paired probabilities introduced by postselection for each node. (e) Fidelity with and without postselection . (f) Postselection ratio and shot count across steps.}
\label{fig:probs_7}
\end{figure}

Analysis of the experimentally obtained results for all three graphs is summarized in Figs.~\ref{fig:probs_4}, \ref{fig:probs_10}, and \ref{fig:probs_7}. The discrepancy between the measured and classically simulated distributions is quantified using the Hellinger fidelity $F_{\rm H}$ and its baseline-corrected version $F_{\rm HBC}$ (see Table~\ref{Tab:graphs} and Methods). As shown in panels (c) in Figs.~\ref{fig:probs_4}, \ref{fig:probs_10}, and \ref{fig:probs_7}, the experimental distributions without postselection are in poor agreement with the ideal distributions. This mismatch is particularly visible at the first steps of the DTQW (panels (d), with statistical errors smaller than the marker size), where high nonuniformity in the probability distribution is expected: the corresponding $F_{\rm H}$ values without postselection are 0.79, 0.61, and 0.57, respectively (see panels (e), with statistical errors smaller than the marker size). This shows that the advanced encoding alone cannot fully preserve coherence.

In contrast, the postselection substantially improves the agreement between the experimental and ideal distributions---the Hellinger fidelity of the resulting probability distributions remains above 0.95 for the 24-qubit circuit and above 0.87 for larger circuits across all DTQW steps. Notably, despite the exponential decrease in the postselection ratio with increasing $t$ (see panels (f)), the symmetry-respecting outcomes remain sufficiently coherent within the endcoding subspace, demonstrating the effectiveness of this noise-suppression approach. The same improvement due to postselection is also visible in the baseline-corrected Hellinger fidelity, which is a more sensitive performance metric in the case the distributions are close to uniform. Overall, these observations confirm that the accuracy of the implemented DTQWs benefits from two complementary approaches: encoding of information within specific symmetry sector of the Hilbert space and applying postselection to preserve this symmetry.

Our approach enables multi-step DTQWs on complex graphs at scales that prior hardware demonstrations had typically not reached, with comparable instances often treated only in simulation~\cite{Nzongani2024-iz, nandi2025robust, sato2024}. Most experimental DTQW implementations to date have relied on alternative encodings and have reported moderate fidelities (e.g., $0.72$--$0.9$), which in practice has limited demonstrations to the 3–6 qubit range and typically 1–4 walk steps~\cite{Acasiete2020-tm, Shakeel2020-hq, Wing-Bocanegra2023, wing2025circuit, razzoli2024efficient}. By contrast, we implement DTQWs using 24–40 qubits on 11–17 node complex graphs while retaining high agreement with the ideal distributions ($F_{\rm H}$ equals $0.95$ and $0.9$, respectively). This comparison indicates that our circuit construction supports substantially larger and less regular graphs on hardware, which is particularly important for biological networks with nontrivial connectivity patterns. 

\subsection*{A case study: Disease gene prioritization}
Discovery of disease pathways typically involves identifying all disease-associated proteins, grouping these proteins into coherent pathways, and analyzing how the resulting pathways connect to the disease at molecular and clinical levels~\cite{agrawal2018}. Network-based methods used for disease protein discovery often operate on the PPI network: given a set of known disease proteins---referred to as seeds---the goal is to predict additional proteins, or targets, that are likely to be associated with the disease. Because many disease pathways are fragmented and weakly embedded within the global interactome, an effective prediction method must be able to operate robustly under such topological constraints~\cite{menche2015uncovering}. DTQWs, which model probabilistic transitions across the network while simultaneously exploring both local and global structures, are naturally suited to this task. 

Here, we demonstrate a proof-of-principle case study, in which DTQW dynamics is able to adequately prioritize asthma-associated nodes within the PPI graph shown in Fig.~\ref{fig:probs_4}. In this graph, the node 7, corresponding to HLA-DQA1 (Major histocompatibility complex, class II, DQ alpha 1), is selected as the starting point for initializing the walker, as it represents prior knowledge about the disease and is one of the most significant asthma-associated genes pertinent to the lowest P‑value in genome-wide association studies (GWAS)~\cite{Clay2022-yp, gkae1070} (HLA-DQA1 lies within the HLA-DQ region, repeatedly implicated in asthma susceptibility through large-scale GWAS and fine-mapping analyses).

\begin{figure}[ht]
\centering
\includegraphics[width=0.9\textwidth]{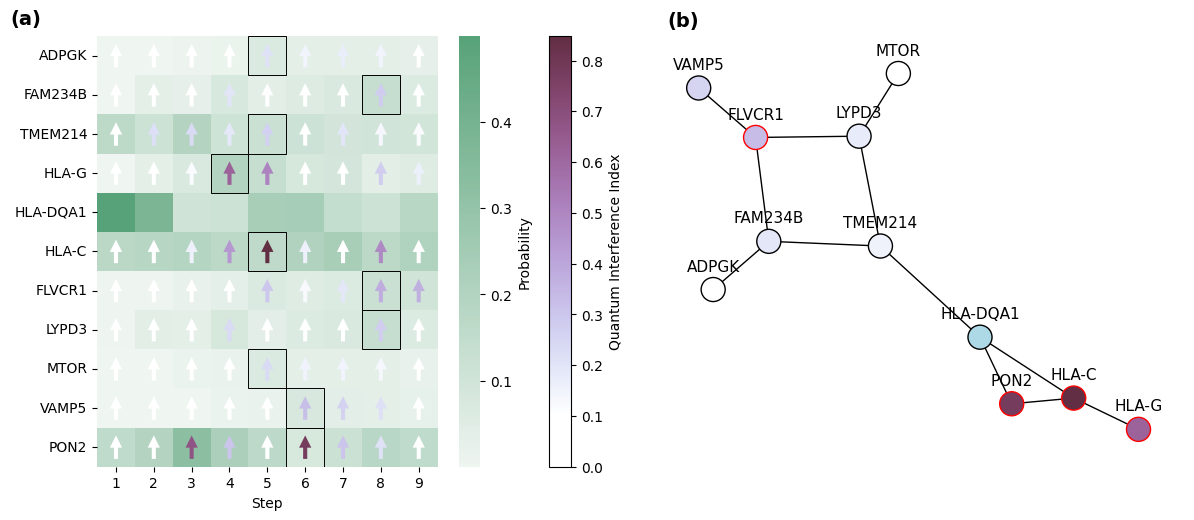} 
\caption{\textbf{Disease-gene prioritization from experimentally measured quantum-walk dynamics.}
\small (a) Heatmap shows postselected transition probabilities in DTQWs on a graph depicted in Fig.~\ref{fig:probs_4}(a). Arrows indicate how strongly quantum interference modifies the probability at node $i$ relative to the classical random walk, and black squares mark the maximum quantum interference index (score). (b) Corresponding biological network with node colors reflecting their scores and the four highest-score genes highlighted in red.} \label{fig:gradient} 
\end{figure}

We expect that interference makes the walker’s probability distribution diverge from the classical one during walker's propagation from the initial node. The nodes, where the divergence is especially visible, are associated with the structural peculiarities of the network. Since the emergence of interference patterns becomes more pronounced only after sufficient propagation depth, we ran two additional experiments beyond the 7-step results, extending the 24-qubit experiment to 9 DTQW steps. As detailed in Methods, we compare the resulting probability distributions with those of a classical random walk to calculate the quantum interference index and assign scores to nodes on its basis, see Fig.~\ref{fig:gradient}. Table~\ref{Tab:prior_genes} reports gene prioritization results ranked according to the DTQW score, with PageRank and DIAMOnD included as benchmark methods and executed using the default parameter settings recommended in their original publications~\cite{PageRank1999, ghiassian2015disease}.

Among the top DTQW‑ranked genes, PON2 (Paraoxonase 2), HLA-C, and HLA-G (Major histocompatibility complex, class I, G) are located near the seed node and were visited early in the walk, consistent with their proximity in the network. In contrast, gene FLVCR1 (Choline and heme transporter 1) with the fourth largest score has no direct graph connections to either the seed node or its neighbors and is assigned lower ranks by the classical benchmark methods. The prioritization of this node in DTQW is attributed to amplitude interference rather than simple topological closeness highlighting DTQW dynamics’ ability to uncover non‑local yet disease-relevant genes. Importantly, FLVCR1 meets the GWAS significance threshold and has been associated with asthma across age groups~\cite{Ferreira2019-ys}.

\begin{table}[ht]
\centering
\caption{Gene prioritization results ranked by the DTQW scores.}\label{Tab:prior_genes}
\begin{tabular}{c c | c c | c | c | l }
\cline{1-7}
 &  & \multicolumn{2}{c|}{\textbf{DTQW}} & \textbf{PageRank} & \textbf{DIAMOnD} & \\
\textbf{Gene} & \textbf{Node} & \textbf{Score} & \textbf{Rank} & \textbf{Rank} & \textbf{Rank} & \textbf{Relation to asthma} \\
\cline{1-7}
HLA-C & 6 & $0.85 \pm 0.012$ & \textbf{1} & \textbf{1} & \textbf{2} & Associated with allergic asthma~\cite{Winnica2022-gv} \\
PON2 & 1 & $0.78 \pm 0.009$ & \textbf{2} & \textbf{2} & \textbf{1} & Therapeutic potential in asthmatics~\cite{Winnica2022-gv} \\
HLA-G & 8 & $0.62 \pm 0.007$ & \textbf{3} & 6 & \textbf{3} & Differentially associated with asthma~\cite{Ribeyre2018-xa} \\
FLVCR1 & 5 & $0.38 \pm 0.006$ & \textbf{4} & 7 & 9 & A risk factor childhood-onset asthma~\cite{Ferreira2019-ys} \\
VAMP5 & 2 & $0.32 \pm 0.007$ & 5 & 10 & 10 & No data \\
LYPD3 & 4 & $0.28 \pm 0.007$ & 6 & 5 & 5 & No data \\
FAM234B & 10 & $0.28 \pm 0.007$ & 7 & \textbf{4} & 7 & No data \\
TMEM214 & 9 & $0.27 \pm 0.008$ & 8 & \textbf{3} & \textbf{4} & No data \\
MTOR & 3 & $0.23 \pm 0.005$ & 9 & 8 & 6 & No data \\
ADPGK & 11 & $0.21 \pm 0.005$ & 10 & 9 & 8 & No data \\
\cline{1-7}
\end{tabular}
\end{table}

The partial overlap between DTQW ranking on one side and PageRank and DIAMOnD rankings on the other side indicates consistency with established approaches, while the difference between them suggests that DTQW can provide complementary information in gene prioritization tasks. The DTQWs applied to the asthma case study operate on an 11-nodes subgraph of the PPI network motif, selected to meet current hardware limitations. While this restricts direct real-world applicability, it serves as a proof of concept for DTQW-based algorithms and illustrates the transition from quantum-inspired algorithms to their actual implementation on quantum hardware. In addition, it demonstrates the advantages of the proposed circuit design.

\subsection*{Scaling analysis}
To assess the scalability of the proposed algorithm, we compare its resource requirements with those of the state-of-the-art approach described by Sato and Saito~\cite{sato2024}. Their quantum circuit design for implementing DTQWs on complex networks requires $\lceil\log_2 N\rceil + \lceil\log_2 |E|\rceil$ qubits. However, this approach relies heavily on multi-controlled Toffoli gates in both shift and coin operators, whose number grows exponentially with the number of qubits involved. Eventually, the coin operator circuit exhibits depth scaling as $\text{poly}(N)$, whereas the depth of shift operator scales as $\text{poly}(|E|)$. This leads to rapidly increasing circuit depth even for relatively small graphs. For instance, a 6-qubit implementation of a single DTQW step on an 8-node, 8-edge graph with maximum degree 3 already decomposes into a circuit containing 1,218 entangling layers, based on the best configuration produced by Qiskit’s \texttt{transpile} function (v1.4.3) at the maximum optimization level without connectivity constraints. When the constraints of heavy-hex qubit connectivity are taken into account, the number of entangling layers increases further, reaching 2,882. The recent paper~\cite{sato2025coinedquantumwalkscomplex} demonstrates that the use of dual-register encoding results in an improved depth scaling, ${\rm poly}(N)$, independent of the number of edges (circuit depth $\propto N^{1.9} \times t$ and exceeds $3 \times 10^3$ for $N=10$ and $t=1$ in the examples considered). Given the limitations of current quantum hardware, deep circuits for $N \geq 9$ (requiring at least $2 \lceil \log_2 N \rceil \geq 8$ qubits) remain challenging for implementation in pre-fault-tolerant quantum computers.

In contrast, in our approach the depth of both the coin and shift operators is independent of $N$ and $|E|$ due to their parallel implementation in qubit space, see Fig.~\ref{fig:implementation}. For the same 8-node, 8-edge example graph as in Ref.~\cite{sato2024}, a single DTQW step generated with our framework requires only 11 entangling layers in QPU with all-to-all connectivity and 28 entangling layers under the heavy-hex connectivity, reducing the circuit depth by two orders of magnitude compared to the aforementioned design. This reduction stems from the scaling properties of our method, in which circuit depth grows primarily with the maximum node degree. Specifically, the depth of the coin operator corresponding to a node with the highest degree $k \equiv \displaystyle \max_i k_i$ does not exceed $3$ for $k = 2$, $11$ for $k = 3$, $35$ for $k = 4$, and $4\lceil \log_2 k \rceil + 8 k - 16$ for $k \geq 5$ in idealized settings without connectivity constraints; $3$ for $k = 2$, $19$ for $k = 3$, $54$ for $k = 4$, and $17 k - 13$ for $k \geq 5$ when heavy-hex connectivity constraints are imposed by hardware. For graphs with a fixed maximum degree, our approach efficiently trades circuit depth for qubit count, reducing the overall circuit volume from $(\lceil\log_2 N\rceil + \lceil\log_2 |E|\rceil) \times (\text{poly}(N) + \text{poly}(|E|)) \times t$ in Ref.~\cite{sato2024} or $2 \lceil\log_2 N\rceil \times \text{poly}(N) \times t$ in Ref.~\cite{sato2025coinedquantumwalkscomplex} down to $2|E| \times {\rm const} \times t$, where the constant scales at most linearly with respect to the highest node degree in the network. Since the average number of gate errors is proportional to the circuit volume, the proposed encoding enables larger-scale experiments on current quantum devices. 

Table~\ref{Tab:graphs} further highlights the scalability of our approach, showing that a 7‑step DTQW circuit on a 17‑node graph requires fewer than 300 entangling layers. In agreement with the scaling analysis, all three graphs in Table~\ref{Tab:graphs}, sharing a maximum degree of 3, result in circuits with similar numbers of entangling layers per step despite differences in graph size and structure. Scaling of the proposed framework for DTQW on complex networks is aligned with the roadmaps of QPU hardware providers~\cite{ibmroad, quantumai, meetiqm} targeting both the substantial growth in the number of physical qubits (up to millions) and the reduction in gate and readout errors as more advanced error-correction codes are being developed.

\section*{Discussion}

In this work, we have presented an implementation strategy for DTQWs on current quantum computers and explored their potential application to gene prioritization in biological networks. The proposed framework has been applied to three model biological graphs, whose corresponding quantum circuits have been successfully executed with reasonable fidelity on current quantum devices at the scale of up to 40 qubits and 287 entangling layers. This feasibility within our method is supported by a combination of the following:
\begin{itemize}
    \item improved gate performance in available hardware, 
    \item encoding information within a specific (single-excitation) symmetry sector of the qubit space, 
    \item circuit optimization unlocked by the sparsity of the used encoding in qubit space,
    \item the use of postselection to filter measurement outcomes respecting the symmetry.
\end{itemize}

Taken together, these features overcome limitations that had previously restricted DTQW experiments to fewer than 10 qubits.
Our proof-of-principle results show the implementability of quantum walks on protein-protein-interaction network motifs with current quantum computers. This illustrates the transition from early quantum-inspired algorithms~\cite{e25050730, Saarinen2024, dubovitskii2025quantum, park2025advancing} to fully hardware-realized quantum algorithms. Our QPU‑experiment‑based node‑prioritization method benefits from quantum interference, thereby providing new information in addition to purely classical methods and showing no contradiction with dedicated biological studies in the asthma case study. Overall, the results highlight both the practical advantages of the proposed circuit design and the requirements posed by existing hardware.

The actual performance of the proposed DTQW algorithm in practice is limited by the average noise level in gates and measurements, as noise leads to the exponential decay of the fraction of symmetry-respecting bitstrings in postselection. Moreover, multiple bit-flip errors may remain undetectable by postselection if the total number of excitations is preserved. These vulnerabilities should be taken into account in the design of larger-scale experiments. Incorporating advanced noise mitigation strategies could improve the experimental results and allow deeper circuits with more steps. For instance, adapting algorithms like the tensor-network error mitigation~\cite{Filippov2023-hc, Fischer2026-vs} for postselection could help reduce error accumulation as circuit depth grows.

As quantum hardware evolves, it will feature more qubits with improved fidelity and connectivity. Although higher fidelity and richer connectivity will enable the reliable execution of longer walk sequences on more complicated graphs, circuit depth will remain a limiting factor. Due to the abundance in physical qubits, whose number is expected to grow up to millions according to the quantum technology roadmaps~\cite{ibmroad, quantumai, meetiqm}, it is much more appealing from a technological perspective to trade the circuit depth for the qubit count in order to exploit the full capability of current hardware. It is the encoding used in our approach that extensively exploits this trade-off, allowing to deal with a wider range of graph structures (of larger sizes and higher densities) as quantum hardware improves. This is a more tractable path forward---compared to boosting fidelities alone---toward practical implementation of complex quantum system dynamics, challenging classical simulations.

While the present study focuses on the scaling behavior of single-particle DTQWs, the underlying framework is inherently extensible to multi-particle, correlated-walker dynamics by incorporating interaction terms within the same hardware-aware circuit design. For example, it paves the way for implementation of the two‑particle DTQW algorithm for distinguishing non‑isomorphic graphs~\cite{berry_two-particle_2011}. Mathematical structure of interacting (entangled) walkers represents an interesting direction for future research, as the dimension of their tensor‑product Hilbert space exponentially increases with the number of walkers and becomes computationally expensive in classical simulations~\cite{Li2025-ry}. In contrast, quantum‑circuit representations can encode such interactions directly through qubit‑qubit coupling, offering a clear advantage. This advantage can find application in multi‑walker node‑prioritization methods applied to heterogeneous biological networks.

\section*{Methods}

\subsection*{Biological data preparation}

We extract gene subgraphs from the BioPlex 3.0 PPI network~\cite{Huttlin2021}, denoted as ${\cal{N}}$. To ensure both structural complexity (the presence of at least two cycles) and biological relevance to asthma, the subgraph generation proceeds in three steps. First, we identify the set ${\cal S}$ of all subgraphs in ${\cal N}$ containing a disjoint 3-cycle and 4-cycle separated by at most two edges. Second, we filter ${\cal S}$ to retain only subgraphs containing at least one asthma-associated gene, based on GWAS summary statistics available through the Open Targets Platform~\cite{gkae1128}. Third, we randomly select a seed subgraph from ${\cal S}$ and iteratively expand it adding neighbors from ${\cal N}$. During each expansion round, we select a node currently in the subgraph with a degree below 3 and randomly pick one of its neighbors in ${\cal N}$. If the neighbor is new to the subgraph, we add both the node and the connecting edge; if the neighbor is already present and its degree is below 3, we add only the edge. This expansion is repeated until the subgraph contains between 12 and 20 edges. This procedure yields a non-tree subgraph with a maximum node degree of 3, carefully balancing topological complexity with the constraints of QPU hardware.

\subsection*{Efficient quantum-walk circuit construction}
DTQW evolution operator for step $t$ reads $U = (CS)^t$ so that each step consists of sequential applications of the shift ($S$) and coin ($C$) operators. The ordering of operators is dictated by the initial state \eqref{eq:state0}, which is an invariant state of the Grover coin operator. The latter is defined on a graph with nodes $i=1,\ldots,N$ through

\begin{equation}
    \label{eq:coin general}
    C = \bigoplus_{i=1}^{N} C_i,
\end{equation}
where $C_i = 2\ket{s_i}\bra{s_i}-I_i$,
\begin{equation}
    \label{eq:superpos}
    \ket{s_i} = \frac{1}{\sqrt{k_i}}\sum_{j:\,(i,j) \in E_{\rm dir}} \ket{i\rightarrow j},
\end{equation}
and $k_i = \sum_{j:\,(i,j) \in E_{\rm dir}} 1$ is the degree of node $i$.

To implement the Grover coin operator \eqref{eq:coin general}, two strategies can be utilized. First, a proposal by Sato et al.~\cite{sato2024} effectively maps the uniform superposition $\ket{s_i}$ to the qubit state $\ket{0\ldots0}$ via the inverse of the initial state preparation process and reflects the state about $\ket{0\ldots0}$ (implements the unitary $2 \ket{0\ldots0} \bra{0\ldots0} - I$, which is equivalent to the multi-controlled Toffoli gate up to the single-qubit gates). Then the reflected state is mapped back to the original coin subspace, preserving the structure of the walker's state \eqref{eq:state}, i.e.,
\begin{equation}
    \label{eq:basis rotate}
    C_i\ket{\psi} = 2\braket{s_i|\psi}\ket{s_i}-\ket{\psi}.
\end{equation}
We find this approach efficient when the nodes have high degree, because it enables a linear scaling of the circuit depth for the coin operator with respect to $k_i \gg 1$. Indeed, a unitary transformation $\ket{0 \ldots 0} \leftrightarrow \ket{s_i}$ is achieved with the circuit of CNOT cost $2 k_i - 2$ and depth $2 \lceil \log_2 k_i \rceil$ in all-to-all connectivity~\cite{cruz-2019} ($2 k_i - 2$ in linear-nearest-neighbor connectivity~\cite{cruz-2019}). For $k_i \geq 5$ the multi-controlled Toffoli gate is implemented without ancilla via a circuit of depth $8 k_i - 16$ and CNOT cost $12 k_i - 44$ in all-to-all connectivity~\cite[Theorem 1, where $n = k_i - 1$]{zindorf-2025} (depth $13 k_i - 9$ and CNOT cost $22 k_i - 62$ in linear-nearest-neighbor connectivity~\cite[Theorem 4, where $k = n+1 = k_i$]{zindorf-2025}). The depth of $C_i$ does not exceed $17k_i - 13$ in linear-nearest-neighbor connectivity and, consequently, heavy-hex connectivity either, ensuring a linear scaling $\sim 17 \displaystyle \max_i\, k_i$ for the depth of the coin operator for large degrees $k_i$.

However, for lower node degrees $k_i \leq 3$ pertinent to the considered protein-protein interaction subgraphs, we propose and utilize a second approach based on embedding the coin operator $C_i$ (acting in the $k_i$-dimensional subspace spanned by vectors $\{ \ket{i \rightarrow j} \}_{j: \, (i,j) \in E_{\rm dir}}$) into the Hilbert space of $k_i$ qubits (with the qubit space dimension being $2^{k_i}$),
\begin{equation}
    \label{eq:embed}
    C_i^{\rm qubits} = I + \sum_{n=1}^{k_i} \sum_{m=1}^{k_i} \left( (C_i^{\rm edges})_{nm} - \delta_{nm} \right) \ket{{\rm bin}(2^{n-1})}\bra{{\rm bin}(2^{m-1})},
\end{equation}
where ${\rm bin}(x)$ denotes the binary representation of integer $x$, with $\ket{{\rm bin}(2^{n-1})}$ being a single-excitation (bracelet) state (one of $\ket{0\dots 001}$, $\ket{0\dots 010}$, $\dots$, $\ket{1\dots 000}$).

\begin{figure}[ht]
\centering
\includegraphics[width=0.3\textwidth]{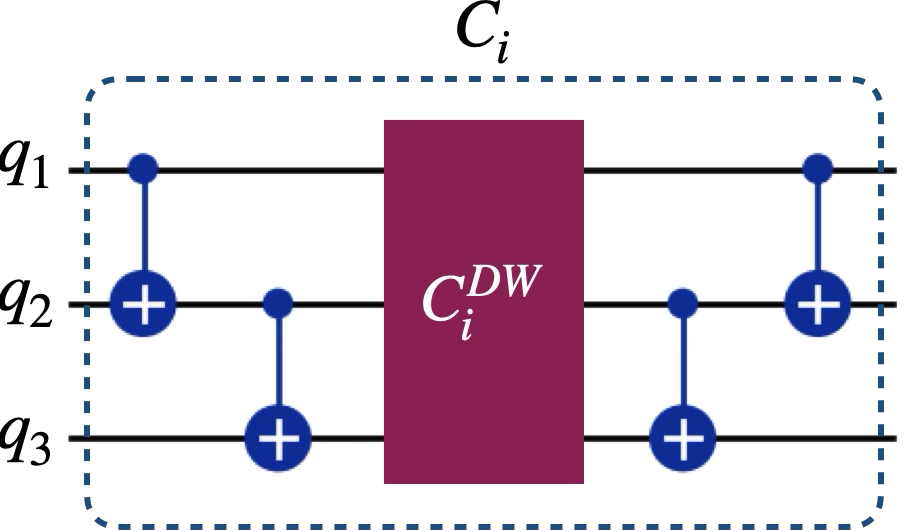} 
\caption{\textbf{Depth-efficient structure of the coin operator ($k_i = 3$).}
\small CNOT cascades map $C_i$ to the operator $C_i^{\rm DW}$ acting on domain-wall states, reducing the overall complexity of decomposition into elementary gates.
} \label{fig:design_b} 
\end{figure}

To optimize the decomposition of multi-qubit gates in $C_i^{\rm qubits}$, we exploit the fact that the Qiskit transpiler can reduce the number of two-qubit operations (and the circuit depth) when the computational basis is rotated~\cite{QiskitUnitariesSynth}. In our case, we permute the bracelet states $\ket{0\dots 001}$, $\ket{0\dots 010}$, $\dots$, $\ket{1\dots 000}$ with the domain-wall (DW) states $\ket{0\dots 001}$, $\ket{0\dots 011}$, $\dots$, $\ket{1\dots 111}$, which is accomplished by the CNOT cascade before and after the unitary operation $C_i^{\rm DW}$, as illustrated in Fig.~\ref{fig:design_b}. $C_i^{\rm DW}$ is given by an analogue of Eq.~\eqref{eq:embed} with the corresponding change in the qubit states. Importantly, the transformation in Fig.~\ref{fig:design_b} temporarily changes the basis in which the Grover coin is expressed, allowing us to significantly reduce the complexity of the decomposed circuit: the number of 2-qubit native entangling gates and the transpiled-circuit depth is approximately halved as compared to those in transpilation of Eq.~\eqref{eq:embed} at the first place. Therefore, the implemented basis rotation reduces entangling operations without altering the logical structure of the algorithm and is particularly useful in pre-fault-tolerant experiments. For example, given the all-to-all (heavy-hex) connectivity, the depth of $C_i$ with $k_i = 3$ is $11$ ($19$) in the latter approach and $14$ ($26$) in the one based on the Toffoli gate and the transformations $\ket{0 \ldots 0} \leftrightarrow \ket{s_i}$, Eq.~\eqref{eq:basis rotate}.

Implementation of the shift operator \eqref{eq:shift} reduces to individual partial swap gates acting on pairs of qubits, see $\alpha$-gates in Fig.~\ref{fig:implementation}(c). The resulting decomposition follows the same sequence of CNOT gates used to realize a full swap, but is augmented with the single-qubit rotation gates, leading to a partial rather than complete swap.

Executing quantum walks on current quantum computers becomes challenging on superconducting hardware with heavy‑hex connectivity, as this topology imposes strong constraints on qubit mapping. The decomposition of shift operators into hardware‑native gates is particularly sensitive to these topological limitations, since a shift operation between non‑adjacent qubits must be transpiled into a sequence of CNOT gates, thereby increasing circuit depth. To mitigate this effect, we transpile each DTQW circuit using Qiskit \texttt{transpile} function 100 times and select the final configuration in the following way. First, we restrict to layouts where the qubits involved in $\alpha$-gates are physically separated by no more than eight edges on the connectivity map. Second, we exclude layouts with ancilla qubits. Third, using the device calibration data retrieved from IBM Quantum Platform~\cite{ibmcloud} prior to circuit submission, we select the layout, in which the cumulative readout error is minimum. Thus, the final layout choice is based on the transpiler output and the hardware calibration data. Altogether, these criteria ensure that the selected layout reduces noise accumulation in the two-qubit gates and measurements, increasing the fraction of measured bitstrings with correct symmetry and reducing the number of undetectable errors. The only purpose of layout optimization is to reduce the effect of noise and support a reasonable level of accuracy in circuits that may include more than 200 entangling layers and measurements on up to 40 qubits.

\subsection*{Experimental setup}

We have performed the experiments on superconducting devices with fixed-frequency transmon qubits as data qubits connected via tunable couplers: the 24- and 40-qubit experiments on IBM Heron R2 QPU \texttt{ibm\_kingston} and the 36-qubit experiment on IBM Heron R3 QPU \texttt{ibm\_pittsburgh}. These devices have 156 qubits, with the median single-qubit gate errors, CZ-gate errors, and readout errors being $2.1 \times 10^{-4}$, $1.7 \times 10^{-3}$, and $4.1 \times 10^{-3}$, respectively, for \texttt{ibm\_pittsburgh} and $2.2 \times 10^{-4}$, $2.1 \times 10^{-3}$, and $8.4 \times 10^{-3}$, respectively, for \texttt{ibm\_kingston} according to the latest calibration prior the experiment. The execution of quantum circuits is performed remotely via IBM Quantum’s cloud infrastructure in Qiskit v1.4.3, an open-source, Python-based, high-performance software stack for quantum computing~\cite{ibmcloud, javadi2024quantum}. 

We set the number of measurement shots to progressively increase with each DTQW step $t$ so as to ensure statistical relevance of the data affected by noise. The number of measurement shots is $5.3 \times 10^5 \times 1.1^{t - 1}$ ($5.3 \times 10^5 \times 1.6^{t - 1}$) in the case of the 24‑qubit (36‑ and 40‑qubit) experiment, which remains below $10^7$ for the steps implemented.

\subsection*{QPU data processing}

The information about the walker location on graph edges $(i,j) \in E_{\rm dir}$ is encoded into the excitation position $q_{(i,j)}$ inside a bitstring $b$ of length $2|E|$, with $b_{q} = \delta_{q,q_{(i,j)}}$. Due to the presence of noise in raw output bitstrings of QPU, the number of excitations is not exactly equal to unity, with no means available to distinguish the noisy excitation from the true one at the level of individual bitstrings. For example, an experimentally observed bitstring $110\ldots 0$ does not belong to the single-excitation subspace, but can be regarded as a noisy version of either $100\ldots 0$ or $010\ldots 0$ from the subspace. The postprocessing of multi-excitation outcomes can be done in several ways, yielding heuristic walker position distributions.

Should each noisy bitstring be replaced by a set of the corresponding single-excitation ones, the total number of the observed excitations is preserved. In this case, the noise-contaminated probability to observe the walker in edge $(i,j) \in E_{\rm dir}$ is therefore given by 
\begin{equation} \label{eq:raw-prob}
P_{(i,j)}^{\rm occ} = \frac{\sum_b b_{q_{(i,j)}}}{\sum_{b,q} b_q}.
\end{equation}
The probability of a walker to be on node $i$ is a marginal probability $P_i^{\rm occ} = \sum_{j:\,(i,j)\in E_{\rm dir}} P_{(i,j)}^{\rm occ}$. The standard error $\Delta P_i^{\rm occ} = \sqrt{P_i^{\rm occ} (1-P_i^{\rm occ}) / \sum_{b,q} b_q}$ does not exceed $4 \times 10^{-4}$ in our experiments. In practice, Eq.~\eqref{eq:raw-prob} corresponds to the average value of the single-qubit operator $\ket{1}\bra{1}$ over qubits, which is further normalized. Eq.~\eqref{eq:raw-prob} can be treated as an occupancy-like quantity, because it is normalized by the total number of detected excitations rather than by the total number of shots.

Each nonzero noisy bitstring $b$ can be replaced by a probabilistic ensemble of the corresponding single-excitation bitstrings, with the lack of precise knowledge about the correct bitstring being reflected in the probability $1/|b|$, where $|b| \equiv \sum_q b_q$ is the Hamming weight of $b$. This way we redistribute multi‑excitation outcomes uniformly over their occupied sites. The zero bitstring can be replaced by any single-excitation bitstring with the uniform probability $1/(2|E|)$, where $2|E|$ is the total number of qubits in our encoding. An effective walker position distribution reads
\begin{equation} \label{eq:ens-prob}
P_{(i,j)}^{\rm eff} = \frac{\sum_{b: \, |b| \neq 0} \frac{1}{|b|} b_{q_{(i,j)}} + \sum_{b: \, |b| = 0} \frac{1}{2|E|} }{\sum_{b} 1}
\end{equation}
and originates from normalization by the total number of shots $\sum_{b} 1$. The probability of a walker to be on node $i$ is a marginal probability $P_i^{\rm eff} = \sum_{j:\,(i,j)\in E_{\rm eff}} P_{(i,j)}^{\rm eff}$. The standard error $\Delta P_i^{\rm eff} = \sqrt{P_i^{\rm eff} (1-P_i^{\rm eff}) / \sum_{b} 1}$ does not exceed $7 \times 10^{-4}$ in our experiments.

In practice, both distributions $P_{(i,j)}^{\rm occ}$ and $P_{(i,j)}^{\rm eff}$ deviate from theoretical one due to the essential noise contamination; however, the latter distribution is in a slightly better agreement with the theory. We therefore use $P_{(i,j)}^{\rm eff}$ as a benchmark for the raw results that are attainable with current quantum computers prone to errors. It also serves as a valuable benchmark for the performance of DTQW algorithms that do not rely on any symmetry verification~\cite{sato2025coinedquantumwalkscomplex}, likewise showing a bad agreement with the theory even at step $t=1$.

Should a quantum computer perform perfectly without errors, then all output bitstrings would have a single excitation, so that Eqs.~\eqref{eq:raw-prob} and \eqref{eq:ens-prob} would both give the same result. Observation of no excitations or more than one excitation is an instance of \emph{detectable} errors. The experimental results substantially improve by removing those detectable errors via postselection. The effect of postselection is to restrict the bitstrings $b$ in Eq.~\eqref{eq:raw-prob} to a single-excitation subspace, the Hamming weight $|b|=1$. This leads to the probability
\begin{equation} \label{eq:ps-prob}
P_{(i,j)}^{\rm ps} = \frac{\sum_{b:\,|b|=1} b_{q_{(i,j)}}}{\sum_{b:\,|b|=1} 1}.
\end{equation}
The probability of a walker to be on node $i$ is a marginal probability $P_i^{\rm ps} = \sum_{j:\,(i,j)\in E_{\rm dir}} P_{(i,j)}^{\rm ps}$. The standard error $\Delta P_i^{\rm ps} = \sqrt{P_i^{\rm ps} (1-P_i^{\rm ps}) / \sum_{b:\,|b|=1} 1}$ does not exceed $3 \times 10^{-3}$ in our experiments. The fraction of bitstrings retained in the postselection procedure equals $(\sum_{b:\,|b|=1} 1) / (\sum_{b} 1)$ and typically decays exponentially with the circuit depth and number of qubits, with the decay rate depending on the gate and measurement error rate of QPU. To make allowance for the decay, we sequentially increase the number of measurement shots with the DTQW step $t$ as reported in Experimental Setup.

It is due to the removal of detectable errors that the probability distribution $P_{(i,j)}^{\rm ps}$ is in good agreement with the theory. However, not all errors are detectable. For example, a state from the encoding subspace does not leave it when affected by phase-flip ($Z$) errors. Similarly, two bitflip errors ($XX$) can convert the state $\ket{10\ldots}$ into $\ket{01\ldots}$ that still belongs to the single-excitation subspace. The rate $\varepsilon_{\rm u.g.}$ of such \emph{undetectable} gate errors determines the scalability of the proposed algorithm. In practice, $\varepsilon_{\rm u.g.}$ is much smaller than the reported errors of individual gates because only some errors are undetectable. The probability of zero undetectable errors can be roughly estimated as $P_0 \approx (1 - \varepsilon_{\rm u.g.})^{N_{\rm gates}(t)} (1-\varepsilon_{\rm u.r.})$, where $N_{\rm gates}(t)$ is the number of two-qubit gates in the circuit depth for step $t$ and $\varepsilon_{\rm u.r.}$ is the undetectable readout error. The latter for $n=2|E|$ qubits in our encoding can be estimated through the single-qubit readout error $\epsilon$ as $\varepsilon_{\rm u.r.} = (n-1)\epsilon^2(1-\epsilon)^{n-2}$, which corresponds to exactly two bitflip errors: one happening in the excited-state qubit and the other in the ground-state qubit. In the experiments performed, $\varepsilon_{\rm u.r.} \sim 5 \times 10^{-4}$. Based on the data in Table~\ref{Tab:graphs} containing the number of gates and the baseline-corrected Hellinger fidelity as an approximation for $P_0$, we estimate $\varepsilon_{\rm u.g.} \sim 3 \times 10^{-4}$. This suggests that a DTQW without postselection would require the gate errors of order approximately $10^{-4}$ in order to reach the same success probability $P_0$ as with active postselection (whereas the currently attainable two-qubit gate errors $\varepsilon_{\rm g}$ are of the order $10^{-3}$).

\subsection*{Performance metrics}

We verify the correctness of the implemented DTQWs by comparing with the classical simulation. To compare the ideal probability distribution $Q$ with the experimental one, $P$, we exploit the Hellinger fidelity $F_{\rm H}(P, Q) = \left( 1-h^2(P, Q) \right)^2 = \left( \sum_i \sqrt{P_i Q_i} \right)^2$, where the Hellinger distance $h(P, Q)$ between $P$ and $Q$ is defined through $h^2(P, Q) = \frac{1}{2} \sum_i \left( {\sqrt{P_i}-\sqrt{Q_i}}\right)^2$. The Hellinger fidelity $F_{\rm H}(P, Q)$ satisfies the property $0\leq F_{\rm H}(P, Q) \leq 1$ and is exactly equal to unity if and only if the two distributions are identical.

In a long-time DTQW experiment ($t \gg 1$), the output of a noisy circuit typically approaches a uniform distribution with increasing circuit depth because unital noise irreversibly increases the entropy and drives the output state to the maximally mixed one~\cite{shtanko-2025} (in practice, local noise could be shaped into the unital form via Pauli twirling~\cite{van_den_berg_probabilistic_2023}). The maximally mixed state corresponds to the uniform distribution $P_{(i,j)}$ over the edge states regardless of the data processing method. The heuristic baseline distribution of the walker over the nodes $i_1, i_2, \ldots$ is therefore $R = (k_{i_1}, k_{i_2}, \ldots) / \sum_j k_j$, where $k_i$ is the degree of node $i$, also known as a stationary or steady-state distribution in the theory of classical random walks on graphs~\cite{lovasz1993random,faccin2013degree}. The effect of noise can therefore be masked by the high values of the Hellinger fidelity $F_{\rm H}(P,Q)$ if $R$ highly overlaps with $Q$. In order to make correction, we consider the baseline-corrected Hellinger fidelity
\begin{equation}
F_{\rm HBC}(P,Q|R) = \frac{F_{\rm H}(P,Q) - F_{\rm H}(R,Q)}{1 - F_{\rm H}(R,Q)}.
\end{equation}
By construction, $F_{\rm HBC}(P,Q|R) \leq 1$ and equals unity if and only if $F_{\rm H}(P,Q) = 1$. If the baseline fidelity $F_{\rm HBC} \leq 0$, then the theoretical distribution $Q$ is in better agreement with the baseline distribution $R$ (null hypothesis) than the experimental distribution $P$. If $P = P_0 Q + (1-P_0)R$, then $F_{\rm H}(P,Q) \approx P_0 + (1-P_0) F_{\rm H}(R,Q)$ and $F_{\rm HBC}(P,Q|R) \approx P_0$.

Uncertainties in $F_{\rm H}$ and $F_{\rm HBC}$ are estimated via Monte Carlo resampling from the empirical distribution. For experiments with active postselection, the uncertainties remain below $10^{-3}$ and $5.5 \times 10^{-3}$, respectively. For experiments without postselection, they remain below $6 \times 10^{-4}$ and $2 \times 10^{-3}$, respectively, across all steps and graphs.

\subsection*{Node prioritization method}
Propagation of a quantum walker through a network is governed by quantum superposition and interference of amplitude states. Unlike classical diffusion processes, the DTQWs do not converge to a stationary distribution; instead, transition probabilities oscillate over time. During these oscillations some nodes $i$ exhibit positive (negative) surges in probability $P_i^{\rm q}(t)$ as compared to the probability $P_i^{\rm cl}(t)$ in classical random walk. The latter is defined through $P_j^{\rm cl}(t+1) = \sum_i M_{ij} P_i^{\rm cl}(t)$, where $M$ is a stochastic matrix defining the Markov chain, $M_{ij} = \alpha \delta_{ij} + (1-\alpha)\delta_{(i,j) \in E_{\rm dir}} k_i^{-1}$. We use the interference quantifier $|P_i^{\rm q}(t) - P_i^{\rm cl}(t)|$ to highlight the nodes with constructive (destructive) quantum interference, thus indicating their functional importance within the network. We exploit the phenomenon of quantum interference to identify the key nodes (i.e., target genes) in biological networks. To quantify this effect, we introduce a quantum interference index (QII)
\begin{equation}
    \label{eq:QII}
    I_i(t) = \frac{|P_i^{\rm q}(t) - P_i^{\rm cl}(t)|}{\sum_j [P_j^{\rm q}(t)]^2}
\end{equation}
in analogy with the measure of quantumness for walks on complex networks~\cite{faccin2013degree}. The collision probability $\sum_j [P_j^{\rm q}(t)]^2 \equiv \braket{P^{\rm q}}$ makes allowance for the average probability of the walker, since the wider is the distribution of the walker, the less is the base for the quantum interference. Importantly, QII vanishes not only for step $t=0$ (initialization of the walker on a seed node) but also for $t=1$ because the initial spreading involves no interference.

A prioritization score is defined as the maximum QII over time,
\begin{equation}
    \label{eq:Score}
    S_i = \max_t I_i(t).
\end{equation}
Since the experimental distribution $P_i^{\rm q}(t)$ approaches the steady-state baseline distribution $R = (k_{i_1}, k_{i_2}, \ldots) / \sum_j k_j$ due to noise and $R = \lim_{t \rightarrow \infty} P^{\rm cl}(t)$~\cite{lovasz1993random,faccin2013degree}, QII vanishes in long-time realistic (noisy) quantum computation. The maximum QII in Eq.~\eqref{eq:Score} is attained within the coherence timescale of QPU qubits. Uncertainties in the prioritization scores are estimated in the same way as for fidelities and are reported in Table~\ref{Tab:prior_genes}.

We associate target genes with the nodes that have the highest scores (exhibit the strongest interference during the DTQWs). This method reveals structural relationships that go beyond simple connectivity and highlights nodes that may be distant from the seed but are nonetheless significant, offering a fundamentally different perspective on biological networks. The method captures both spatial and temporal aspects of DTQW dynamics, effectively marking the distinctive nodes in walker's propagation through the network. 

\section*{Data availability}

Data is available from the corresponding author upon reasonable request.

\section*{Code availability}
The design code with details of the DTQW quantum circuit construction is openly available at 

\noindent \href{https://doi.org/10.6084/m9.figshare.32218227}{https://doi.org/10.6084/m9.figshare.32218227}

\section*{Acknowledgements}
This work was conducted within the framework of the Healthcare and Life Sciences Quantum Working Group. The authors would like to thank Guillermo Garc\'{\i}a-P\'{e}rez for useful conversations regarding this work. No funding was received for this research.
% Acknowledge any financial support or technical assistance here.

\section*{Author contributions}

V.D., S.F., F.U., A.B., and L.P. conceptualized the study; S.F. conceived the experiment; A.B. performed data preprocessing; S.F., V.D., and F.U. performed the experiments; V.D. processed the results and drafted the original version of the manuscript; V.D., F.U., A.B., and S.F. interpreted the results and wrote the manuscript; L.P. and S.M. supervised the work; all authors participated in the discussion of the results. All authors have read and approved the manuscript.

\section*{Competing interests}
The authors declare no competing financial or non-financial interests.

%% --- References ---
\bibliography{bibliography} % Your .bib file

\end{document}